\documentclass[12pt,draftclsnofoot,onecolumn]{IEEEtran}
\usepackage{amssymb}
\usepackage{amsfonts}
\usepackage{amsfonts,subfigure,multicol,color,verbatim,
graphicx,cite,epsfig,amssymb,amsmath,cases,bm,algorithm,
algorithmic,xcolor,multirow,array}

\begin{document}
\markboth{Submit to IEEE Wireless Communications, vol. x, no. y,
Mon. 2014} {Peng: C-RANs \ldots}

\title{Fronthaul-Constrained Cloud Radio Access
Networks: Insights and Challenges}

\author{
Mugen~Peng, \IEEEmembership{Senior Member,~IEEE},
Chonggang~Wang,~\IEEEmembership{Senior Member,~IEEE},
Vincent~Lau,~\IEEEmembership{Fellow,~IEEE}, and H. Vincent
Poor,~\IEEEmembership{Fellow,~IEEE}
\thanks{Mugen~Peng (e-mail: {\tt pmg@bupt.edu.cn}) is
with the Key Laboratory of Universal Wireless Communications for
Ministry of Education, Beijing University of Posts and
Telecommunications, Beijing, China. Chonggang~Wang (e-mail: {\tt
cgwang@ieee.org}) is with InterDigital Communications, King of
Prussia, PA, USA. Vincent Lau (e-mail: {\tt eeknlau@ee.ust.hk}) is
with the Dept. of ECE, Hong Kong University of Science and
Technology, Hong Kong. H. Vincent Poor (e-mail: {\tt
poor@princeton.edu}) is with the School of Engineering and Applied
Science, Princeton University, Princeton, NJ, USA. }
\thanks{The work of M. Peng was supported in part by the National Natural Science Foundation of
China under Grant 61222103 and 61361166005, the National Basic
Research Program of China (973 Program) under Grant 2013CB336600,
and the Beijing Natural Science Foundation under Grant 4131003. The
work of V. Lau was supported by N\_HKUST605/13. The work of H. V.
Poor was supported in part by the U.S. National Science Foundation
under Grant ECCS-1343210.} }

\date{\today}
\renewcommand{\baselinestretch}{1.5}
\thispagestyle{empty} \maketitle \thispagestyle{empty}
\vspace{-15mm}
\begin{abstract}
As a promising paradigm for fifth generation (5G) wireless
communication systems, cloud radio access networks (C-RANs) have
been shown to reduce both capital and operating expenditures, as
well as to provide high spectral efficiency (SE) and energy
efficiency (EE). The fronthaul in such networks, defined as the
transmission link between a baseband unit (BBU) and a remote radio
head (RRH), requires high capacity, but is often constrained. This
article comprehensively surveys recent advances in
fronthaul-constrained C-RANs, including
 system architectures and key techniques. In particular, key
techniques for alleviating the impact of constrained fronthaul on
SE/EE and quality of service for users, including compression and
quantization, large-scale coordinated processing and clustering, and
resource allocation optimization, are discussed. Open issues in
terms of software-defined networking, network function
virtualization, and partial centralization are also identified.
\end{abstract}


\newpage

\section{Introduction}

The mobile communication industry is currently developing the fifth
generation (5G) system, with the objective of providing pervasive
always-on, always-connected broadband packet services. It is widely
agreed that compared to the fourth generation (4G) system, 5G should
achieve system capacity growth by a factor of 1000, and spectral
efficiency (SE), energy efficiency (EE) and data rate growth all by
a factor of 10 \textcolor[rgb]{1.00,0.00,0.00}{\cite{I:DP}}. To
achieve these goals, radically new technologies need to be
developed. Inspired by the green soft cloud/collaborative/clean
access networks in\textcolor[rgb]{1.00,0.00,0.00}{\cite{I:5G}}, the
cloud radio access network (C-RAN) has been proposed as a
combination of emerging technologies from both the wireless and the
information technology (IT) industries by incorporating cloud
computing into radio access networks (RANs).

\begin{figure}[!h]
\centering  \vspace*{0pt}
\includegraphics[scale=0.45]{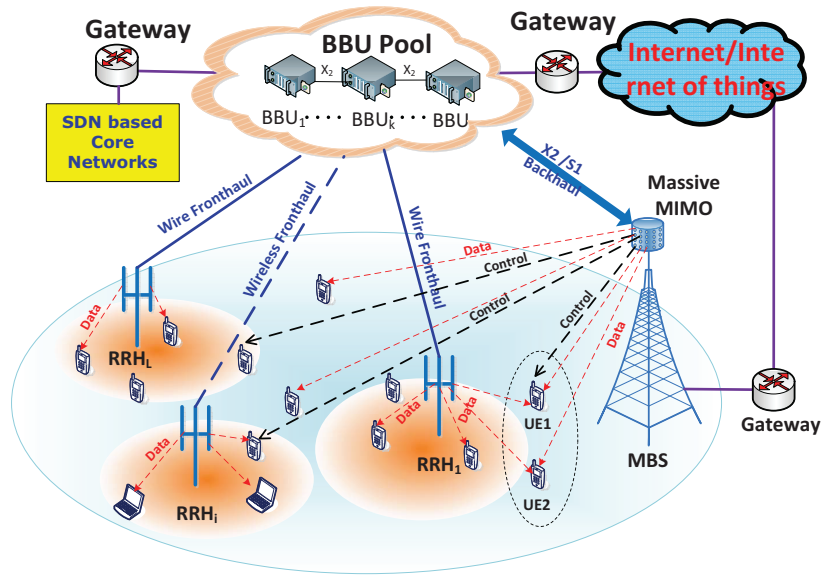}
\setlength{\belowcaptionskip}{-100pt} \vspace*{-10pt} \caption{The
components and system architecture of C-RANs} \vspace*{-10pt}
\label{CRAN}
\end{figure}

In C-RANs, the traditional base station (BS) functions are decoupled
into two parts: the distributed installed remote radio heads (RRHs)
and the baseband units (BBUs) clustered as a BBU pool in a
centralized cloud server. RRHs support seamless coverage and provide
high capacity in hot spots, while BBUs provide the large-scale
processing and management of signals transmitted/received from
diverse RRHs, where the cloud computing technology provides flexible
spectrum management and advanced network coordination. Through the
fronthaul between RRHs and BBUs, the cost-effective and
power-efficient RRHs can operate as soft relays by compressing and
forwarding the signals received from user equipments (UEs) to the
centralized BBU pool, as depicted in Fig. \ref{CRAN}. Further, the
system can be developed using software radio technology to further
enable all centralized processing to use platforms based on an open
IT architecture, which makes upgrading to different RAN standards
possible without hardware upgrades. Additional advantages of C-RANs
include advanced technology facilitation, resource
virtualization/cloudization, low energy consumption, and efficient
interference mitigation through large-scale cooperative processing.

%paragraph: states

Since the C-RAN architecture was first proposed by China Mobile in
2011\textcolor[rgb]{1.00,0.00,0.00}{\cite{I:5G}}, further research
and development has been pursued. With the C-RAN architecture,
mobile operators can quickly deploy RRHs to expand and make upgrades
to their networks. Therefore, the C-RAN has been advocated by both
operators (e.g., France Telecom/Orange, Telefonica, and China
Mobile) and equipment vendors (e.g., Alcatel-Lucent LightRadio, and
Nokia-Siemens LiquidRadio) as a means to achieve the significant
performance gains required for 5G. At the same time, several C-RAN
projects have been initiated in many organizations such as the Next
Generation Mobile Networks (NGMN) project, the European Commission's
Seventh Framework Programme, etc.

Despite their attractive advantages, C-RANs also come with their own
challenges in connection with the fronthaul links created by this
architecture. As presented in Fig. \ref{CRAN}, a prerequisite
requirement for the centralized processing in the BBU pool is an
inter-connection fronthaul with high bandwidth and low latency.
Unfortunately, practical fronthaul is technology capacity or
time-delay constrained, which has a significant impact on SE and EE
performance gains of C-RANs. To overcome the disadvantages of C-RANs
imposed by the fronthaul constraints, compression and large-scale
pre-coding/de-coding with low overhead are required. Additionally,
radio resource allocation optimization taking the constrained
fronthaul into account is also key to mitigating interference,
guaranteeing the diverse quality of service (QoS) requirements for
different UEs, and achieving significant SE and EE performance
gains.

%paragraph:consider the abstract, address the contributions and difference of this work
A number of studies of feasible designs and operations of C-RANs
have been published recently.
In\textcolor[rgb]{1.00,0.00,0.00}{\cite{HCRAN}}, the heterogeneous
cloud radio access network (H-CRAN) as a 5G paradigm toward green
and soft themes is briefly presented to enhance C-RAN. To alleviate
the capacity constraint on the fronthaul links, a multi-service
small-cell wireless access architecture based on combining
radio-over-fiber with optical wavelength division multiplexing (WDM)
techniques is proposed
in\textcolor[rgb]{1.00,0.00,0.00}{\cite{I:CLiu}}. Unfortunately,
until now the characteristics of fronthaul-constrained C-RANs have
not been highlighted, and a framework for improving SE/EE
performance and guaranteeing diverse QoS requirements for different
UEs with reference to the system architecture and key techniques has
not been treated in depth. In this article, we present a
comprehensive survey of technological features and core principles
of fronthaul-constrained C-RANs. In particular, the system
architecture of C-RANs is presented, and key techniques including
large-scale coordinated processing and clustering, compression and
quantization, and resource allocation optimization to improve SE/EE
performance and guarantee diverse QoS requirements in
fronthaul-constrained C-RANs are summarized. Challenging open issues
related to fronthaul-constrained C-RANs, including software-defined
networking (SDN), network function virtualization (NFV), and partial
centralization, are discussed as well.

The remainder of this paper is organized as follows. C-RAN system
architectures are introduced in Section II. Compression and
quantization techniques to improve SE and EE performance for
fronthaul-constrained C-RANs are presented in Section III.
Large-scale coordinated processing and clustering techniques are
introduced in Section IV. Resource allocation optimization
techniques adaptive to diverse and time-varying packet services are
discussed in Section V. Future challenges are highlighted in Section
VI, followed by conclusions in Section VII.

\section{C-RAN System Architectures}

A C-RAN centralizes different baseband processing resources to form
a single resource pool, such that the resource can be managed and
dynamically allocated on demand. C-RANs have several advantages over
traditional cellular  architectures, such as increased resource
utilization efficiency, low energy consumption, and light
interference.

\subsection{C-RAN Components}

The general architecture of a C-RAN consists of three components,
namely (i) a BBU pool consisting of a large number of BBUs with
centralized processors, (ii) RRHs with antennas located at remote
sites, and (iii) a fronthaul network that connects RRHs to BBUs with
high capacity and low time latency.

\subsubsection{RRH}

RRHs are mainly used to provide high data rate for UEs with basic
wireless signal coverage, by transmitting radio frequency (RF)
signals to UEs in the downlink and forwarding the baseband signals
from UEs to the BBU pool for centralized processing in the uplink.
In general, RRHs perform RF amplification, up/down conversion,
filtering, analog-to-digital conversion, digital-to-analog
conversion, and interface adaptation. By conducting most signal
processing functions in the BBU pool, RRHs can be relatively simple,
and can be distributed in a large-scale scenario with a
cost-efficient manner.

\subsubsection{BBU Pool}

A BBU pool is located at a centralized site and consists of
time-varying sets of software defined BBUs, which operate as virtual
BSs to process baseband signals and optimize radio resource
allocation. In the software defined BBU, the signal processing
resources are dynamically allocated and the processing capability is
adaptively reconfigured based on traffic-aware scheduling of UEs and
time-varying radio channels. The radio resources of different BBUs
can be fully shared, and thus a large-scale virtual multiple-input
multiple-output (MIMO) system is formed from the BBU pool's
perspective.

\subsubsection{Fronthaul}

Fronthaul is defined as the link between BBUs and RRHs, and its
typical protocols include the common public radio interface (CPRI)
and the open base station architecture initiative
(OBSAI)\textcolor[rgb]{1.00,0.00,0.00}{\cite{II:CPRI}}. Fronthaul
can be realized via different technologies, such as optical fiber
communication, cellular communication, and even millimeter wave
communication. Generally, fronthaul falls into two categories: ideal
without any constraints, and non-ideal with bandwidth, time latency
and jitter constraints. Optical fiber communication without
constraints is considered to be the ideal fronthaul for C-RANs
because it can provide high transmission capacity at the expense of
high cost and inflexible deployment. By contrast with optical fiber,
wireless fronthauls employing cellular or microwave communication
technologies with carrier frequencies between 5 and 40 Gigahertz
(GHz) are cheaper and more flexible to deploy, at the expense of
capacity and other constraints. Since wireless fronthaul or
capacity-constrained optical fiber is cheap and flexible, these
technologies are anticipated to be prominent in practical C-RANs,
and thus this article will focus only on such non-ideal constrained
fronthaul.

\subsection{C-RAN System Structures}

\begin{figure}[!h]
\centering  \vspace*{0pt}
\includegraphics[scale=0.36]{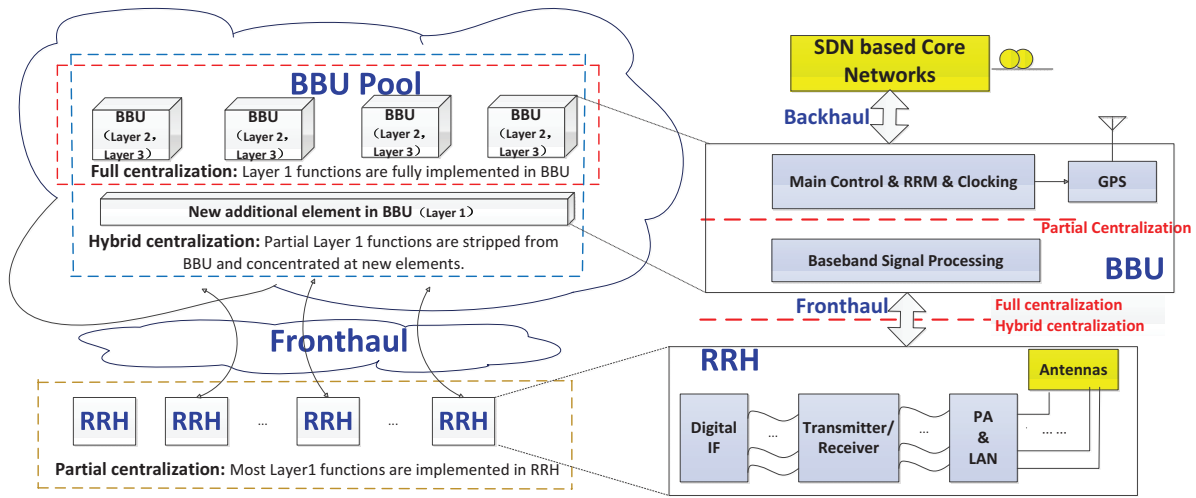}
\setlength{\belowcaptionskip}{-100pt} \vspace*{-10pt} \caption{Three
options of C-RAN system structures} \vspace*{-10pt}
\label{CRAN_architecture}
\end{figure}

Although there are various possibilities for C-RAN structures,
according to the constraints on fronthaul and the distribution of
functions between BBUs and RRHs, three options are categorized as
shown in Fig. \ref{CRAN_architecture}.

\textbf{Full centralization}: This is also called a stacked BBU
structure\textcolor[rgb]{1.00,0.00,0.00}{\cite{I:5G}}, where
functions of the baseband (i.e., physical layer, Layer 1), the
medium access control layer (MAC, Layer 2), and the network layer
(Layer 3) of the conventional BS are moved into the BBU. This option
is the premier C-RAN configuration, which is clear and simple, but
incurs a high burden on fronthaul. The BBU contains all processing
and managing functions of the traditional BS. Since this structure
has significant benefits in terms of operation and maintenance,
significant attention has been devoted to developing techniques to
alleviate the heavy burden on the fronthaul of this system.

\textbf{Partial centralization}: This is also called the standalone
Layer 1 structure, in which the RRH integrates not only the RF
functions but also some RF related baseband processing functions,
while all other functions in Layer 1 and the upper layers are still
located in the BBU. This option greatly reduces the RRH-BBU overhead
and alleviates the constraints on fronthaul since Layer 1 bears the
major computational burden of RANs. However, some advanced features
such as coordinated multiple point transmission and reception (CoMP)
and spatial cooperative processing for distributed massive MIMO
cannot be efficiently supported. The interaction between Layer 2 and
Layer 1 can also be complex, which increases the difficulty of
interconnection between Layer 2 and Layer 1. In other words, this
structure is still far from being practical.

\textbf{Hybrid centralization}: This is regarded as a special case
of full centralization, in which partial functions in Layer 1 such
as the user specific or cell specific signal processing functions
are removed from BBUs, and assembled into a separate processing
unit, which may be a part of the BBU pool. The benefit of this
structure is its flexibility to support resource sharing and the
potential capability to reduce modifications and energy consumption
in BBUs.

For the fully centralized structure, the performance of C-RANs is
clearly constrained by the fronthaul link capacity. In the uplink,
for instance, RRHs need to sample, quantize, and then forward the
received RF signals to the BBU pool. With densely deployed RRHs, the
fronthaul traffic generated from a single UE with several MHz
bandwidth could be easily scaled up to multiple gigabits per second
(Gbps). In practice, a commercial fiber link with tens of GHz
capacity could thus be easily overwhelmed even under moderate mobile
traffic. One approach to this problem is to utilize the partially
centralization structure though substantial signal processing
capabilities are required on RRHs; the other alternative is to adopt
advanced techniques to optimize the performance under a fully
centralized structure with constrained fronthaul. To simplify
functions and capabilities of RRHs, the latter solution is the focus
of this article, and the corresponding key techniques are surveyed.

\section{Signal Compression and Quantization}

A C-RAN can be regarded as a special case of a relay structure with
a wireless first-hop link and a wireless/fiber second-hop link.
Signal compression/quantization is critical to alleviating the
impact of fronthaul constraints on SE and EE performance. From an
information theoretic perspective, the effects of compression and
quantization can be modeled via a test channel, with the
uncompressed signals as the input and the compressed signals as the
output. The test channel is often modeled as a Gaussian channel for
simplicity of analysis, in which the output signal is generated from
the input signal corrupted by an additive Gaussian compression
noise. The design of a codebook for this channel is equivalent to
setting the variance of the compression noise. An interesting result
shows that by simply setting the quantization noise power
proportional to the background noise level at each RRH, the
quantize-and-forward scheme can achieve a capacity within a constant
gap to a throughput performance upper (i.e., cutset)
bound\textcolor[rgb]{1.00,0.00,0.00}{\cite{II:compression}}.

\subsection{Compression and Quantization in the Uplink}

In the uplink (UL), each RRH compresses its received signal and
forwards the compressed data to the central BBU pool as a soft relay
through the limited-capacity fronthaul link. The central BBU pool
then performs joint decoding of all UEs based on all received
compressed signals. Compared to conventional independent compression
across RRHs, distributed source coding strategies are generally
beneficial since signals received at different RRHs are
statistically
correlated\textcolor[rgb]{1.00,0.00,0.00}{\cite{III:SHPar}}. By
leveraging signals received from other RRHs as side information,
distributed source coding can reduce the rate of the compressed
stream by introducing some uncertainty into the compressed signal
that is resolvable and enables the quality of the compressed signal
received from the desired RRH to be improved. The amount of rate
reduction that is allowed without incurring decompression errors
depends critically on the quality of the side information, which
should be known to the encoder. When multiple RRHs compress and
forward their received signals to the BBU pool, the compression
design is transformed into the problem of setting the covariance
matrix of the compression noises across different RRHs. In this
setting, distributed Wyner-Ziv lossy compression can be used at RRHs
to exploit the signal correlation. In particular, considering the
fact that the observations of all coordinating RRHs are actually
correlated because they are broadcast from the same source,
distributed Wyner-Ziv compression can be applied to exploit this
correlation and reduce the required fronthaul transmission rate.

However, the distributed compression techniques require each RRH to
have information about the joint statistics of the received signals
across all RRHs, and they are generally sensitive to uncertainties
regarding the side information. Therefore, the implementation of
distributed Wyner-Ziv compression is difficult mainly due to the
high complexity of determining the optimal joint compression
codebook and the joint decompressing/decoding at the BBU pool.
Alternatively, for the independent compression method, the
quantization codebook of an RRH is determined only by its local
channel state information (CSI), and the decompression operation at
the BBU pool is also on a per-RRH basis. Although independent
quantization reduces the fronthaul complexity compared to joint
compression design, the compression codebook generation is still
based on information theoretic source coding techniques, which can
be highly complex and impractical in a rapidly varying wireless
environment. A robust compression scheme for a practical scenario
with inaccurate statistical information about the correlation among
the RRHs' signals was proposed
in\textcolor[rgb]{1.00,0.00,0.00}{\cite{III:SHPar}}. By using an
additive error model with bounds on eigenvalues of the error matrix
to model the inaccuracy, the problem is formulated and a
corresponding solution is provided to achieve a stationary point for
the problem by solving Karush-Kuhn-Tucker conditions. It is observed
that, despite the imperfect statistical information, the proposed
robust compression scheme can tolerate sizable errors with drastic
performance degradation while maintaining the benefits of
distributed source coding.

\begin{figure}[!h]
\centering  \vspace*{0pt}
\includegraphics[scale=0.6]{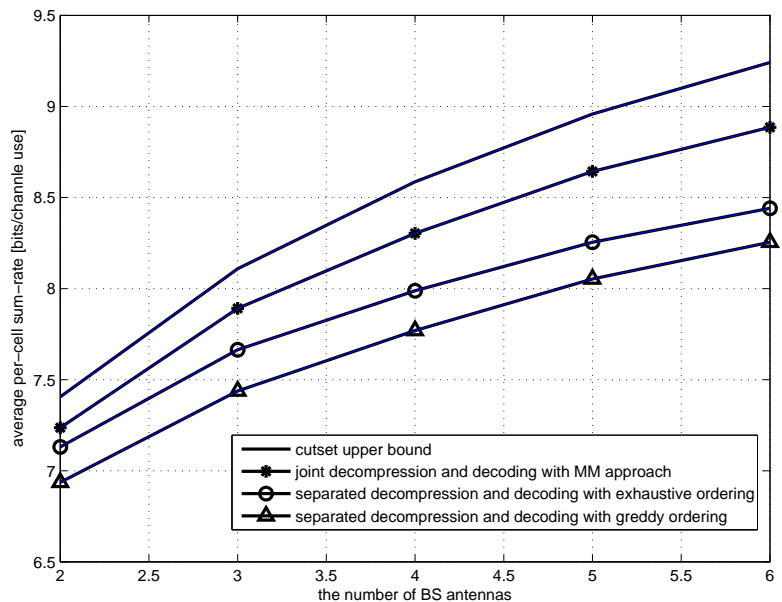}
\setlength{\belowcaptionskip}{-100pt} \vspace*{-10pt}
\caption{Performance comparisons between the joint and separated
decompression and decoding approaches, where the number of cells is
assumed at 3, the UE transmission power is 20 dBm, the fronthaul
capacity is constrained by 12 bit/c.u., and the inter-cell channel
gain is assumed to be -10
dB\textcolor[rgb]{1.00,0.00,0.00}{\cite{III:SPJon}}.}
\vspace*{-10pt} \label{JDD2}
\end{figure}

It is worth noting that a potential advantageous approach for C-RANs
is when the decoder performs joint decompression and decoding. This
approach was first studied
in\textcolor[rgb]{1.00,0.00,0.00}{\cite{III:SPJon}} for the scenario
with multi-antenna BSs and multi-antenna UEs. The sum-rate
maximization problem with joint decompression and decoding under the
assumption of Gaussian test channels is shown to be an instance of
non-convex optimization. To solve this problem, an iterative
algorithm based on the majorization minimization (MM) approach was
proposed to guarantee the convergence to a stationary point of the
sum-rate maximization. As shown in Fig. \ref{JDD2}, the advantage of
the proposed joint decompression and decoding with MM has been
demonstrated compared to the conventional approaches based on the
separate decompression and decoding with exhaustive ordering or with
greedy ordering. Further, the rates achieved by this approach are
much closer to the cutset upper bound than those of separate
decompression and decoding approaches.

\subsection{Compression and Quantization in the Downlink}

Similarly to the UL, the compression-and-forward strategy is also
useful in the downlink (DL), where the BBU pool first pre-codes each
message for UEs to allow for interference collaboration both across
UEs and among data streams for the same UE, and then compresses the
pre-coded signals for the distributed UEs via RRHs. As a counterpart
of the distributed source coding strategy for the UL, the joint
design of precoding and compression based on multivariate
compression for finite-capacity DL fronthaul links was studied
in\textcolor[rgb]{1.00,0.00,0.00}{\cite{III:SHPJo}}. Unlike
conventional point-to-point compression, the effects of the additive
quantization noises at UEs can be controlled through proper design
of the correlation of the quantization noises across RRHs. The
problem of maximizing the weighted sum-rate subject to the fronthaul
constraints over the precoding matrix and the compression noise
covariance for given weights can be formulated and solved. It is
confirmed in\textcolor[rgb]{1.00,0.00,0.00}{\cite{III:SHPJo}} that
the joint precoding and compression strategy outperforms
conventional approaches based on the separate design of precoding
and compression or independent compression across RRHs, especially
when the transmit power or the inter-cell channel gain are large, or
when the limitation imposed by the finite-capacity fronthaul links
is significant. Notably, it is observed that multi-terminal
compression strategies provide performance gains of more than 60\%
for both UL and DL in terms of the cell-edge throughput.

Instead of the aforementioned pure compression strategy with
constrained fronthaul, a hybrid compression and message-sharing
strategy for DL transmission is presented
in\textcolor[rgb]{1.00,0.00,0.00}{\cite{III:PPatil}}. In the
proposed strategy, the BBU pool directly sends messages for some UEs
to the RRH along with the compressed pre-coded signals of the
remaining UEs. An overall algorithm to optimize the hybrid strategy
involving the choice of beamforming vector, power, quantization
noise levels, and more importantly the decision of which users
should participate in the direct message transmission and which
users in compression is presented. The numerical results show that
the hybrid strategy can achieve a saving in fronthaul capacity of
about 60\% compared with the message-direct transmission scheme, and
improve the rate of the 50th percentile user by about 10\% at the
same fronthaul capability compared with the pure compression scheme,
which demonstrates that this approach outperforms the pure
compression and the pure message-direct transmission schemes.

\section{Coordinated
signal processing and Clustering}

Large-scale coordinated signal processing in the BBU pool is
considered one of the most promising techniques to improve network
capacity performance. It is necessary to exploit precoding and
decoding coefficient design schemes that make special considerations
for the capacity-constrained fronthaul. Full-scale coordination in a
large-scale C-RAN requires the processing of very large channel
matrices, leading to high computational complexity and channel
estimation overhead. For example, when the optimal linear receiver
is adopted, the computational complexity typically grows cubically
with the precoding matrix size. This implies that the average
computational complexity per RRH or per UE grows quadratically with
the matrix size, which fundamentally limits the scale of RRH
cooperation. One potential solution is to decompose the overall
channel matrix into small sub-matrices, which results in adopting
the user-centric clustering technique. According to the clustering
technique, a sub-matrix can be formed and processed separately,
although this would inevitably cause performance loss.

\subsection{Precoding Techniques}

There are two types of in-phase and quadrature (IQ)-data transfer
methods in fronthaul: 1) \textbf{after-precoding}, with which a BBU
transfers IQ-data after precoding data symbols with a beamforming
matrix or vector, and 2) \textbf{before-precoding}, with which a BBU
transfers beamforming weights for each data stream and data symbols
separately before data symbols are precoded. The required bit-rate
for after-precoding IQ-data transfer depends only on the number
antennas used for transmission/reception at the RRH. With
after-precoding, all the information for IQ-data should be exchanged
for each symbol between the BBU and the RRH. In contrast, with
before-precoding IQ-data transfer, data symbols for each user are
exchanged for each symbol duration, but beamforming weights for each
data stream are exchanged less frequently according to the channel
coherence time.

In a C-RAN, only a small fraction of the overall entries in the
channel matrix have reasonably large gains, since any given user is
only close to a small number of RRHs in its neighborhood, and vice
versa. Thus, ignoring the small entries in the channel matrix can
significantly sparsify the matrix, which potentially leads to a
great reduction in processing complexity and channel estimation
overhead. Therefore, group sparsity is often required rather than
individual sparsity in practical C-RANs. For example, the mixed
$l_1/l_p$ -norm is adopted
in\textcolor[rgb]{1.00,0.00,0.00}{\cite{IV:pre10}} to induce
sparsity in large-scale cooperative C-RANs. Two group sparse
beamforming (GSBF) algorithms of different complexities are
proposed: namely, a bi-section GSBF algorithm and an iterative GSBF
algorithm. It is demonstrated that the GSBF framework is very
effective in providing a near-optimal solution. The bi-section GSBF
algorithm proves to be a better option for large scale C-RANs due to
its low complexity. The iterative GSBF algorithm can be applied to
provide better performance in a medium-size network. Note that all
the precoding techniques mentioned in this section assume the
availability of perfect CSI at the BBU. It is therefore apparent
that the acquisition of perfect CSI is critical to optimal precoding
design in C-RANs; however, obtaining perfect CSI is very challenging
because many parameters are involved, leading to significant
estimation errors, quantization errors and feedback delays.

\subsection{Clustering Techniques}

By controlling the number of RRHs in one cluster for a typical user,
clustering techniques can limit the estimation overhead and
computational complexity to a low level, which results in low
capacity requirements on the fronthaul. However, this technique
inevitably reduces the C-RAN large-scale cooperative processing
gains and lowers the C-RAN capacity. By limiting the scale of
coordinated RRHs within a small cluster, the centralized processing
capability is not fully exploited.

There are two types of RRH clustering schemes: disjoint clustering
and user-centric clustering. In disjoint clustering, the entire
C-RAN is divided into non-overlapping clusters and RRHs in each
cluster jointly serve all UEs within the coverage area. Although
disjoint clustering has already shown its effectiveness in
mitigating the inter-cell interference, those UEs at the cluster
edge still suffer from considerable inter-cluster interference.
Alternatively, in user-centric clustering, each UE is served by an
individually selected subset of neighboring RRHs and different
clusters for different UEss may overlap. The benefit of user-centric
clustering is that there exists no explicit cluster edge. The
user-centric clustering scheme can be further categorized into two
different implementations depending on whether the RRH clustering is
dynamic or static over different time slots. In dynamic user-centric
clustering, the RRH cluster for each UE can change over time,
allowing for more freedom to fully utilize the fronthaul resources.
However, dynamic user-centric clustering also requires a large
amount of signaling overhead as new UE associations need to be
established continuously. In static user-centric clustering, the UE
association is fixed over time and may only need to be updated as
the UE location changes. Dynamic clustering can significantly
outperform the disjoint clustering strategy, while the heuristic
static clustering schemes can achieve a substantial portion of the
performance gain.

The cluster size is a critical system design parameter. Optimal
dimensioning of the cluster size is necessitated by the facts that:
1) the cluster size determines the number of active RRHs at any
given time. In turn the density of active RRHs shapes the co-channel
interference experienced by a scheduled UE; 2) the radius of the
cluster characterizes the number of concurrently scheduled UEs per
unit area; and 3) the dimensions of a cluster determines the number
of RRHs serving a scheduled UE. This in turn determines the
diversity gain experienced due to spatially distributed RRHs.

\begin{figure}[t!]
\centering
\includegraphics[width=3.5in]{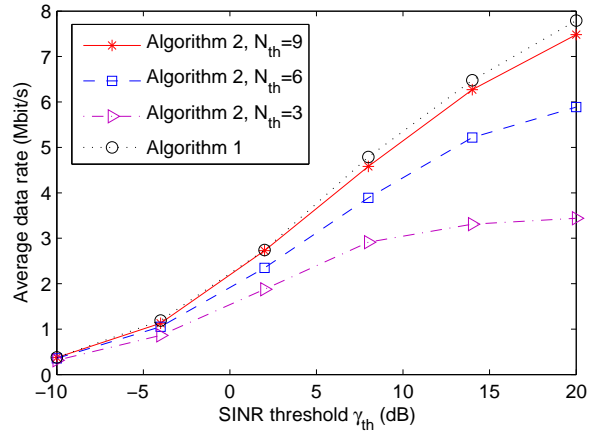}
\caption{Average data rate comparisons between the dynamical
clustering solution (Algorithm 2) with limited fronthaul and
Algorithm 1 with ideal fronthaul, where the frequency bandwidth is
assumed to be 5 MHz, the radius of the effective region is set at
500 m, the transmit power is set at 23 dBm, the transmit power of
interfering RRH nodes is set at 23 dBm, the noise power spectral
density is set at -162 dBm/Hz, and the path loss exponent is set at
4.} \label{fig:clus size cons}
\end{figure}

In\textcolor[rgb]{1.00,0.00,0.00}{\cite{zhaozhongyuan}}, an explicit
expression for the successful access probability (SAP) for clustered
RRHs is derived by applying stochastic geometry. By using the
obtained theoretical result as a utility function, the clustering of
RRHs is formulated as a coalitional formation game model, and a
merge and split method is proposed as an efficient solution. Fig.
\ref{fig:clus size cons} provides the average data rate of the
proposed solution (named Algorithm 2), where the fronthaul
constraint is considered. The maximum number of RRHs in each C-RAN
cluster is set as $N_{\mathrm{th}} = 3, 6, 9$, which means that the
maximum cluster size per user is constrained. The average data rate
increases as $N_{\mathrm{th}}$ increases. By allowing the formation
of larger clusters, more interference can be removed, and thus the
performance can be improved. Moreover, the average data rate
approaches that of Algorithm 1 without fronthaul constraints when
the threshold of cluster size is set as $N_{\mathrm{th}} = 9$.
Compared with Algorithm 1, the cluster size is restricted by the
given threshold. The simulation results show that the distribution
of Algorithm 2 is almost coincident with that of Algorithm 1 when
the value of $N_{\mathrm{th}}$ is large enough.

\section{Radio Resource Allocation and Optimization}

In fronthaul-constrained C-RANs, more advanced radio resource
allocation and optimization (RRAO) techniques are required than in
traditional cellular networks due to the densely distributed RRHs
and the powerful centralized BBUs. Multi-dimensional joint resource
optimization including precoding, resource block (RB) allocation,
user scheduling, power allocation, and cell association can
significantly enhance overall system performance and maintain
satisfactory QoS for UEs in C-RANs. Due to the large-scale
cooperative processing among RRHs and the delay-tolerant packet
traffic arriving in an unpredictable and bursty fashion, RRAO comes
with extraordinary challenges because the problem is largely
intractable due to its non-convex nature.

To deal with the delay-aware RRAO problem, there are mainly three
approaches, including equivalent rate constraint, Lyapunov
optimization, and Markov decision processes (MDPs). The equivalent
rate constraint approach is to convert the average time delay
constraints into equivalent average rate constraints using queuing
theory or large deviations theory. The Lyapunov optimization
approach is to convert the average delay constraints into a
minimization of the Lyapunov drift-plus-utility function. The MDP
approach is a systematic approach to dealing with delay-aware RRAO
by solving the derived Bellman equation in a stochastic learning or
differential equation setting. Compared to the equivalent rate
constraint approach and Lyapunov optimization approach, MDP can
achieve the best performance at the expense of the highest
complexity.

In\textcolor[rgb]{1.00,0.00,0.00}{\cite{Lijian}}, a hybrid
coordinated multi-point transmission (H-CoMP) scheme is presented
for downlink transmission in fronthaul constrained C-RANs, which
fulfills the flexible tradeoff between large-scale cooperation
processing gain and fronthaul consumption. H-CoMP splits the traffic
payload into shared streams and private streams. By reconstructing
the shared streams and private streams to optimize pre-coders and
de-correlators, the shared streams and private streams can be
simultaneously transmitted to obtain the maximum achievable degrees
of freedom (DoF) under the limited fronthaul constraints. To
minimize the transmission delay of the delay-sensitive traffic under
the average power and fronthaul consumption constraints in C-RANs,
the queue-aware rate and power allocation problem is formulated as
an infinite horizon average cost constrained partially observed
Markov decision process (POMDP). The queue-aware H-CoMP (QAH-CoMP)
solution adaptive to both the urgent queue state information (QSI)
and the imperfect channel state information at transmitters (CSIT)
in the downlink C-RANs is obtained by solving a per-stage
optimization for the observed system state at each scheduling frame.
Since QAH-CoMP requires centralized implementation and perfect
knowledge of CSIT statistics, and has exponential complexity with
respect to (w.r.t.) the number of UEs, the linear approximation of
post-decision value functions involving POMDP is presented. Further,
a stochastic gradient is proposed to allocate power and transmission
rate dynamically with low computing complexity and high robustness
against the variations and uncertainties caused by unpredictable
random traffic arrivals and imperfect CSIT. Furthermore, online
learning is used to estimate the per-queue post-decision value
functions and update the Lagrange multipliers effectively.

\begin{figure}[t!]
\centering
\includegraphics[width=4.5in]{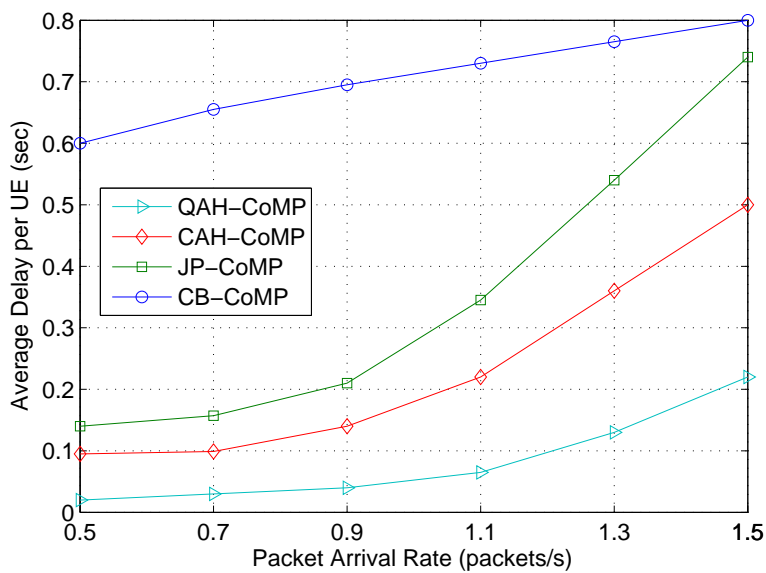}
\caption{Average packet delay vs. packet arrival rate, the maximum
capacity fronthaul constraint is assumed to be 20 Mbit/s, the
maximum transmit power is 10 dBm, the mean size of traffic packet is
4 Mbit/s, the maximum buffer size is 32 Mbits, the total frequency
bandwidth is 20 MHz, and the scheduling frame duration is 10 ms.}
\label{fig:MAP}
\end{figure}

To compare performance gains of QAH-CoMP, three baselines are
considered in simulations: coordinated beamforming (CB)-CoMP, joint
processing (JP)-CoMP, and channel-aware resource allocation with
H-CoMP (CAH-CoMP). All these three baselines carry out rate and
power allocation to maximize the average system throughput with the
same fronthaul capacity and average power consumption constraints as
the proposed QAH-CoMP. For the CB-CoMP baseline, the BBU pool
calculates the coordinated beamformer for each RRH to eliminate the
dominating intra-cluster interference among RRHs. For JP-CoMP, all
RRHs are jointly coordinated to suppress the RRH-interference in the
BBU pool. For the CAH-CoMP baseline, H-CoMP transmission is adopted,
while the power allocation and rate allocation are only adaptive to
CSIT. Fig. \ref{fig:MAP} compares the delay performance of these
four solutions for different packet arrival rates. The average
packet delay of all the schemes increases as the average packet
arrival rate increases. Compared with CB-CoMP, JP-ComP can provide
better delay performance. It is noted that the delay performance
gains of JP-ComP diminish as the packet arrival rate increases,
which is due to the fact that the fronthaul capacity becomes
relatively limited with the increasing packet arrival rate. Both
CAH-CoMP and QAH-CoMP can provide better performance than the
traditional CB-CoMP and JP-CoMP due to the contribution of H-CoMP.
Apparently, the performance gain of QAH-CoMP compared with CAH-CoMP
is contributed by power and rate allocation with the consideration
of both urgent traffic flows and imperfect CSIT.

\section{Challenging Work and Open Issues}

Although there has been some progress on the above mentioned
potential system architectures and advanced techniques for fronthaul
constrained C-RANs, there are still many challenges ahead, including
C-RANs with SDN, C-RANs with NFV, partially centralized C-RANs,
standards development, field trials, etc. Since the C-RAN standards
developments in 3GPP are still not open, and the corresponding field
trials for 5G systems are far in the future, this section will
discuss the aforementioned first three challenges.

\subsection{C-RANs with SDN}

SDN is an emerging architecture that decouples the network control
and forwarding functions, which enables the network control to
become directly programmable and the underlying infrastructure to be
abstracted to adjust applications and network services dynamically.
The network intelligence is logically centralized in the SDN which
maintains a global view of the network, which in turn enables network
managers to configure, manage, secure, and optimize resources
quickly via dynamic, automated SDN programs. As one key component in
5G, SDN will significantly influence the C-RAN interfaces to the
core network.

SDN enables the separation of the control plane and the data plane,
which can also be applied to C-RAN environments. However, currently
envisioned C-RANs do not support the control and data decoupling
functions. Rather, C-RANs have mainly been considered as means
presented to provide data transmission with high bit rates. Due to
high density RRHs, too much signaling overhead for the system
control and broadcasting information will be necessitated, which
will greatly degrade the C-RAN performance. Therefore, a new system
architecture design for the control plane is needed to be compliant
with the SDN architecture. Fortunately, heterogeneous cloud radio
access networks (H-CRANs) have been proposed to efficiently support
the control and data decoupling functions by incorporating macro
base stations (MBSs) in C-RANs, where MBSs are used to deliver the
control signaling and provide seamless
coverage\textcolor[rgb]{1.00,0.00,0.00}{\cite{HCRAN}}. Therefore,
how to design the interface between H-CRANs and the SDN based core
network is an issue that should be investigated further. Since
C-RANs allow the aggregation of traditional base station resources,
and the SDN allows for load sharing, finding ways of combining them
effectively is an important topic for future research.

\subsection{C-RANs with NFV}

NFV refers to the implementation of network functions in software
running on general purpose computing/storage platforms. This
approach allows the deployment of network functions in data centers
and leveraging of traffic load through virtualization techniques. By
contrast, the state of the art is to implement network functions on
dedicated and application specific hardware. Hence, the main
motivation for NFV is to exploit the economy of high volume hardware
platforms, to reduce life and innovation cycles within
telecommunication networks through software updates rather than
hardware updates, and to exploit novel data center technologyies.
The C-RAN's core feature is resource cloudization, in which the
centralized resources can be dynamically allocated to form a soft
BBU entity. Given current vendors' proprietary and closed platforms,
it is advantageous to develop an NFV based BBU platform for modern
data centers, which consolidate many network equipment types onto
industry standard high volume servers, switches and storage, and
which could be located in data centers, network nodes and in end
user premises\textcolor[rgb]{1.00,0.00,0.00}{\cite{II:NFV}}.

The virtualization of the BBU pool in C-RANs through NFV is still
not straightforward. When incorporating NFV into C-RANs, the BBU
pool should be deployed on multiple standard servers, in which there
is an additional dedicated hardware accelerator for the
computation-intensive physical layer function processes. Meanwhile,
the additional hardware accelerator design should be required to
meet the strict real-time requirements for wireless signal
processing. The functionalities in upper layers as well as
additional user applications, such as content distribution and web
caching, could be implemented through a virtual machine. It is
anticipated that by applying the NFV principles, most of the current
radio signal processing and networking could be implemented in a
general purpose computing environment, allowing new virtualized
functions to be added to the network dynamically and intelligently.

\subsection{C-RANs with Inter-Connected RRHs}

To alleviate constraints of capacity and time latency on the
fronthaul of C-RANs, distributed cooperation among inter-connected
RRHs with the partial centralization structure is a promising
alternative solution.
In\textcolor[rgb]{1.00,0.00,0.00}{\cite{Fogcomputing}}, fog
computing is proposed to provide computing, storage, and networking
services between end devices and traditional cloud computing data
centers, typically, but not exclusively located at the edge of
network, which can be used in C-RANs to alleviate the constraints of
fronthaul and high computing capabilities in the BBU pool through
transferring some cooperative processing into RRHs and even users
from the centralized BBU pool. Therefore, interconnections between
some RRHs are possible and the corresponding topology should be
investigated.
In\textcolor[rgb]{1.00,0.00,0.00}{\cite{ConnectedRRH}}, both mesh
and tree-like backhaul network architectures are presented to
decrease the negative influences of capacity-constrained backhaul
links on CoMP. The trade-off between wireless cluster feasibility
and the backhaul connectivity level are designed carefully.
Particularly, compared to the mesh backhaul network architecture,
the wireless cluster feasibility in the tree topology is about 50
percent lower with significant reduced network deployment and
maintenance costs. Inspired by this backhaul clustering technique,
the mesh and tree-like fronthaul clustering can be utilized to
balance cluster feasibility and connectivity in partial
centralization structure based C-RANs.

Based on the partial centralization structure of fog computing based
C-RANs, the inter-RRH interference can be suppressed in the
centralized BBU pool with a high capacity constraint on fronthaul,
or can be handled collaboratively between the distributed and
adjacent RRHs without any constraints on fronthaul. Therefore, the
topology should be adaptively configured to balance the constraints
of fronthaul and the complexity of distribution cooperative
processing. The advanced sparse beamforming and dynamic RRH
clustering technique should be jointly optimized. In addition, the
information asymmetry between the BBU pool and the connected RRH
should be highlighted, and a contract based game model could be
effective to optimize the radio resource allocation.

\section{Conclusions}

This article has outlined and surveyed the state-of-the-art system
architecture design, key techniques and future work for fronthaul
constrained C-RANs. With the goal of understanding further
intricacies of key techniques, we have broadly divided the body of
knowledge into signal compression and quantization, coordinated
signal processing and clustering, and radio resource allocation
optimization. Within each of these aspects, we have summarized the
diverse problems and the corresponding solutions that have been
proposed. Nevertheless, given the relative infancy of the field,
there are still quite a number of outstanding problems that need
further investigation. Notably, it is concluded that greater
attention should be focused on transforming the C-RAN paradigm into
an SDN and NVF framework with fog computing.

\begin{IEEEbiography}{Mugen Peng}
(M'05--SM'11) received the B.E. degree in Electronics Engineering
from Nanjing University of Posts \& Telecommunications, China in
2000 and a PhD degree in Communication and Information System from
the Beijing University of Posts \& Telecommunications (BUPT), China
in 2005. After the PhD graduation, he joined in BUPT, and has become
a full professor with the school of information and communication
engineering in BUPT since Oct. 2012. During 2014, he is also an
academic visiting fellow in Princeton University, USA. He is leading
a research group focusing on wireless transmission and networking
technologies in the Key Laboratory of Universal Wireless
Communications (Ministry of Education) at BUPT. His main research
areas include wireless communication theory, radio signal processing
and convex optimizations, with particular interests in cooperative
communication, radio network coding, self-organizing network,
heterogeneous network, and cloud communication. He has
authored/coauthored over 50 refereed IEEE journal papers and over
200 conference proceeding papers.

Dr. Peng is currently on the Editorial/Associate Editorial Board of
the \emph{IEEE Communications Magazine}, the IEEE Access, the
\emph{IET Communications}, the \emph{International Journal of
Antennas and Propagation} (IJAP), the \emph{China Communications},
and the \emph{International Journal of Communications System}
(IJCS). He has been the guest leading editor for the special issues
in the \emph{IEEE Wireless Communications}. Dr. Peng was a recipient
of the 2014 IEEE ComSoc AP Outstanding Young Researcher Award, and
the best paper award in GameNets 2014, CIT 2014, ICCTA 2011, IC-BNMT
2010, and IET CCWMC 2009. He received the First Grade Award of
Technological Invention Award in Ministry of Education of China for
his excellent research work on the hierarchical cooperative
communication theory and technologies, and the Second Grade Award of
Scientific and Technical Advancement from China Institute of
Communications for his excellent research work on the co-existence
of multi-radio access networks and the 3G spectrum management.
\end{IEEEbiography}

\begin{IEEEbiography}{Chonggang Wang}
(SM'09) received the B.S. degree from Northwestern Polytechnical
University (NPU), Xi'an, China, in 1996, the M.S. degree from the
University of Electronic Science and Technology of China (UESTC),
Chengdu, China, in 1999, and the Ph.D. degree from the Beijing
University of Posts and Telecommunications (BUPT), Beijing, China,
in 2002. He is currently a Member Technical Staff with the
Innovation Lab, InterDigital Communications, King of Prussia, PA,
USA, with a focus on Internet of Things (IoT) R\&D activities
including technology development and standardization. His current
research interest includes IoT, machine-to-machine (M2M)
communications, mobile and cloud computing, and big data management
and analytics. Dr. Wang was a corecipient of the National Award for
Science and Technology Achievement in Telecommunications in 2004 on
IP Quality of Service (QoS) from the China Institute of
Communications. He is the founding Editor-in-Chief of IEEE Internet
of Things Journal, an Advisory Board Member of The Institute-IEEE
(2015-2017), and on the editorial board for several journals
including IEEE Transactions on Big Data and IEEE Access. He has been
selected as an IEEE ComSoc Distinguished Lecturer (2015-2016).
\end{IEEEbiography}

\begin{IEEEbiography}{Vincent Lau}
(F'12) obtained B.Eng (Distinction 1st Hons) from the University of
Hong Kong (1989-1992) and Ph.D. from the Cambridge University
(1995-1997). He joined Bell Labs from 1997-2004 and the Department
of ECE, Hong Kong University of Science and Technology (HKUST) in
2004. He is currently a Chair Professor and the Founding Director of
Huawei-HKUST Joint Innovation Lab at HKUST.  His current research
focus includes robust cross layer optimization for MIMO/OFDM
wireless systems, interference mitigation techniques for wireless
networks, device-to-device communications as well as networked
control systems.
\end{IEEEbiography}

\begin{IEEEbiography}{H. Vincent Poor}
(S'72, M'77, SM'82, F'87) received the Ph.D. degree in EECS from
Princeton University in 1977.  From 1977 until 1990, he was on the
faculty of the University of Illinois at Urbana-Champaign. Since
1990 he has been on the faculty at Princeton, where he is the
Michael Henry Strater University Professor of Electrical Engineering
and Dean of the School of Engineering and Applied Science. Dr.
Poor's research interests are in the areas of stochastic analysis,
statistical signal processing, and information theory, and their
applications in wireless networks and related fields such as social
networks and smart grid. Among his publications in these areas are
the recent books \textit{Principles of Cognitive Radio} (Cambridge
University Press, 2013) and \textit{Mechanisms and Games for Dynamic
Spectrum Allocation} (Cambridge University Press, 2014).

Dr. Poor is a member of the National Academy of Engineering and the
National Academy of Sciences, and a foreign member of Academia
Europaea and the Royal Society. He is also a fellow of the American
Academy of Arts and Sciences, the Royal Academy of Engineering
(U.K.), and the Royal Society of Edinburgh. He received the Marconi
and Armstrong Awards of the IEEE Communications Society in 2007 and
2009, respectively. Recent recognition of his work includes the 2014
URSI Booker Gold Medal, and honorary doctorates from several
universities in Europe and Asia, including an honorary D.Sc. from
Aalto University in 2014.
\end{IEEEbiography}

\end{document}